\documentclass[%
superscriptaddress,
 amsmath,amssymb,
prx,
twocolumn,
]{revtex4-1}
\usepackage[export]{adjustbox}
\usepackage{graphicx}
\usepackage{color}
\usepackage{xcolor}
\usepackage{wasysym}
\usepackage{dcolumn}
\usepackage{bm}
\usepackage{hyperref}
\usepackage[mathscr]{euscript}
\graphicspath{{fig/}}
\definecolor{mygreen}{rgb}{0,0.5,0}
\definecolor{mybrown}{rgb}{0.65,0.16,0.16}
\newcommand{\colg}[1]{\textcolor{mygreen}{#1}}

\newcommand{\colb}[1]{\textcolor{blue}{#1}}
\newcommand{\colr}[1]{\textcolor{red}{#1}}
\newcommand{\colm}[1]{\textcolor{magenta}{#1}}

\newcommand{\colc}[1]{\textcolor{cyan}{#1}}
\newcommand{\colbr}[1]{\textcolor{mybrown}{#1}}
\newcommand{\colo}[1]{\textcolor{orange}{#1}}

\def\mbf {\mathbf}
\def\gr{\Gamma_{r}}
\def\rel{R_{\lambda}}
\def\nr{r/\eta}
\def\la{\langle}
\def\ra{\rangle}
\def\epsm{\la{\epsilon}\ra}
\def\gcube{{\gr^3/r^4\epsm}}
\def\delx{\Delta x}
\def\beq {\begin{equation}}
\def\eeq {\end{equation}}
\def\uu{\mbf{u}}
\def\uo{\mbf{\omega}}
\def\un{\mbf{n}}

\def\ul{\mbf{l}}
\def\dho{\partial}
\def\dur{\Delta_r u}
\def\dvr{\Delta_r v}

\def\dlllnr{\la (\Delta_r u)^3 \ra/r\epsm}
\def\ar{{\alpha (r)}}
\def\zp{{\lambda_p}}
\def\zpm{{\lambda_{|p|}}}

\def\nabv{\mbf\nabla}

\def\circr{\Gamma_r}
\def\beqa {\begin{eqnarray}}
\def\eeqa {\end{eqnarray}}

\def\meandiss {\langle\epsilon\rangle}

\begin{document}
\title{Circulation in high Reynolds number isotropic turbulence is a bifractal}

\author{Kartik P. Iyer}
\affiliation{Department of Mechanical Engineering, New York University, New York, NY 11201, USA}
\author{Katepalli R. Sreenivasan} 
\email{krs3@nyu.edu}
\affiliation{
Department of Mechanical Engineering, New York University, 
New York, NY, $11201$, USA
}
\affiliation{
Department of Physics and the Courant Institute of Mathematical Sciences, New York University, New York,
NY $11201$, USA
}
\author{P. K. Yeung}
\affiliation{
Schools of Aerospace Engineering and Mechanical Engineering, Georgia Institute of Technology, Atlanta, GA $30332$, USA
}

\date{\today}


\begin{abstract}
The turbulence problem at the level of scaling exponents is hard in part because of the multifractal scaling of small scales, which demands that each moment order be treated and understood independently. This conclusion derives from studies of velocity structure functions, energy dissipation, enstrophy density (that is, square of vorticity), etc. However, it is likely that there exist other physically pertinent quantities with uncomplicated structure in the inertial range, potentially resulting in huge simplifications in the turbulence theory. We show that velocity circulation around closed loops is such a quantity. By using a large databases of isotropic turbulence, generated from numerical simulations of the Navier-Stokes equations over a wide range of Reynolds numbers, we show that circulation exhibits a bifractal behavior at the highest Reynolds number considered: space filling for moments up to order $3$ and a mono-fractal with an unchanging dimension of about $2.5$ for higher orders; this change in character roughly at the third-order moment is reminiscent of a ``phase transition". We explore the possibility that circulation becomes effectively space filling at much higher Reynolds numbers even though it may technically be regarded as a bifractal. We confirm that the circulation properties depend on only the area of the loop, not its shape; and, for a figure-$8$ loop, the relevant area is the scalar sum of the two segments of the loop. 
\end{abstract}

\maketitle
From Leonardo da Vinci's half-a-millennium old drawings of turbulent motions in the river Arno \cite{vinci} to their visualizations on modern day computers \cite{kaneda03}, evidence abounds that turbulent motion comprises organized structures, often evocatively described as ``eddies" and ``vortices". On the other hand, both phenomenological and analytical theories of turbulence \cite{K41a,MY.II} have largely focused on multi-point correlators of velocity, whose connection to the physical structures is not always clear. Furthermore, the most obvious multi-point correlators are best described as multifractals \cite{Benzi1984,Fri95,Eyink1995,SA97}, which makes the problem very difficult to explore analytically. In this paper, we show that the velocity circulation around closed loops, besides providing a plausible link between vortical structures and statistical objects, has a very simple bifractal structure: space filling for moments roughly up to order $3$ and a mono-fractal with an unchanging dimension of about $2.5$ for higher orders. We comment on possible Reynolds number effects. 

For reference, the circulation around a loop of linear dimension $r$ is defined as
\beq
\label{def.eq}
\gr \equiv \oint_{\dho D_r} \uu(\ul)\cdot d\ul = \oiint_{D_r} \uo \cdot {\hat{\un}}dA  \;,
\eeq
where $\dho D_r$ denotes the boundary of the loop, $\uu$ is the velocity, $\uo \equiv \nabv \times \uu$ is the vorticity, $d\ul$ is an elemental length along $\dho D_r$ and ${\hat{\un}}dA$ is an elemental area of $D_r$. Migdal \cite{migdal}, who initiated the statistical theory of circulation, made the case that, in the inertial range that is far from both the forcing and dissipation scales, the probability density function (PDF) of $\gr$ depends uniquely on the scaling variable $G = \gr^{2k}/r^{4k-2}$, if the loop size $r$ lies within the inertial range, where $k$ is some parameter. Kolmogorov's arguments (henceforth K41) \cite{K41a} lead to $k=3/2$, with the circulation moments scaling as $\la \gr^p \ra \sim r^{4p/3}$. Migdal also argued that the PDF depends only on the area circumscribed by a simply-connected loop, and not on its actual shape, as long as it is entirely contained in the inertial range. This is the area rule. Another result (call it the figure-$8$ area rule) is that, when the loop is a figure-$8$ with a common vertex, the circulation statistics depend only on the scalar sum of the areas of the component segments of figure-$8$, rather than their vectorial sum as one traverses the loop. 

This theory was followed up soon after by a small number of experimental and numerical papers \cite{umeki,KRSJS95,CSKRS96,benzi97,zhou08}. Because they were all limited to low Reynolds numbers, the verification of the area rules was stymied by the modest extent of the inertial range; and, obviously, they could not evaluate the high-Reynolds-number properties. Further, the experimental flows were not homogeneous, which thus complicated the inferences drawn therefrom. It is thus not a great surprise that these earlier studies did not agree quantitatively among themselves. One common inference of these early studies is that circulation is highly intermittent, just as the velocity increments are, and display multifractal scaling with no unique value of $k$, making no new insights possible. We assess these properties persuasively here by taking recourse to direct numerical simulation (DNS) data of statistically stationary, homogeneous and isotropic turbulence in a periodic box over a wide range of Reynolds numbers \cite{pkpnas}, including the highest of any so far \citep{YSP18}. We also show that the scaling variable $G = \gr^3/r^4$, corresponding to $k=3/2$ in Migdal's expression \cite{migdal}, stemming from K41, shows a very good collapse in the inertial range. 

Migdal did not have conclusive thoughts on the scaling of circulation moments. We show that circulation at the highest Reynolds numbers of this study is a bifractal, to which we have already made an allusion; this simple behavior contrasts a highly intermittent, multifractal structure typical of low Reynolds numbers. This result goes a considerable distance in addressing the question: What statistical variable in turbulence would be most apt to study, among the infinite number of possibilities?

For later purposes, we provide the following definitions. The longitudinal velocity increment is defined as $\dur \equiv u(x+r)-u(x)$, where the velocity component $u$ and the separation distance $r$ are both taken in the same direction. The inertial range is defined as that range where the normalized third-order velocity structure function $\dlllnr$, $\langle \epsilon \rangle$ being the global mean value of the energy dissipation rate, is equal to the exact theoretical value of $-4/5$ \cite{K41b}. This result has been evaluated in Ref.~\cite{KI16}, for the same data as the present.

\section{Data}
The DNS data used in this work have been acquired by solving the incompressible Navier-Stokes equations,
\beq
\partial \mathbf{u}/\partial t
+(\mathbf{u} \cdot \nabla) \mathbf{u}
= -\nabla (p/\rho) + \nu \nabla^2 \mathbf{u} + \mathbf{f} \;,
\eeq
where $\mathbf{u}$ is the solenoidal velocity field ($\nabla \cdot \mathbf{u}=0$), $p$ is pressure, $\rho$ is fluid density, $\nu$ is the kinematic viscosity and $\mathbf{f}$ is the forcing term that maintains a stationary state \citep{EP88,DY10}. We use Fourier pseudo-spectral calculations \citep{rogallo} on a periodic domain of size $(2\pi)^3$ with an explicit second order Runge-Kutta integration in time. A combination of phase-shifting and truncation is used to reduce aliasing errors, where the highest resolved wavenumber $k_{max}=\sqrt{2}N/3$ and $N$ is the number of grid points in one direction. Typical spatial resolution, expressed by $k_{max}\eta$, was fixed around $1.5$ in earlier simulations \citep{IGK2009}. Recently \citep{YSP18}, it has been pointed out that the spatial and temporal resolution are more stringent at higher Reynolds numbers. For some of the data analyzed here, data were obtained with improved resolution. Table \ref{dns.tab} lists some flow parameters of interest. 

\begin{table}
\centering
\caption{Isotropic DNS database: $N^3$ is the number of points on a $L_0^3$ grid with $L_0 = 2\pi$ units, $\rel \equiv u'\lambda/\nu$ is the Taylor scale Reynolds number where $u'$ is the root-mean-square velocity fluctuation,
$\lambda \equiv u'/\sqrt{\la(\dho u/\dho x)^2\ra}$ is the Taylor microscale, $\nu$ is the kinematic viscosity, $L \approx L_0/5$ is the integral scale, $\eta \equiv (\nu^3/\epsm)^{1/4}$ is the Kolmogorov scale, $\delx = L_0/N$ is the grid spacing. Results have been averaged over a time span of at least $10$ large-eddy time scales $(L/u')$, except for the $16,384^3$ data for which the averaging time is much shorter.}
\label{dns.tab}

\begin{tabular}{ccccc}
\\
\hline
$N^3$ & $\rel$ & $L/\eta$ & $\delx/\eta$ & $\epsm L/u'^3$ \\
\hline
$256^3$ & $140$ & $108$ & $2.1$ & $0.44$\\
$512^3$ & $240$ & $226$ & $2.1$ & $0.42$\\
$2048^3$ & $400$ & $446$ & $1.1$ & $0.41$ \\
$4096^3$ & $650$ & $898$ & $1.1$ & $0.39$\\
$8192^3$ & $1300$ & $2514$ & $1.5$ & $0.38$\\
$16384^3$ & $1300$ & $2522$ & $0.8$ & $0.39$\\
\hline
\end{tabular}
\end{table}
\begin{figure}
\centering
\includegraphics[width=0.99\linewidth]{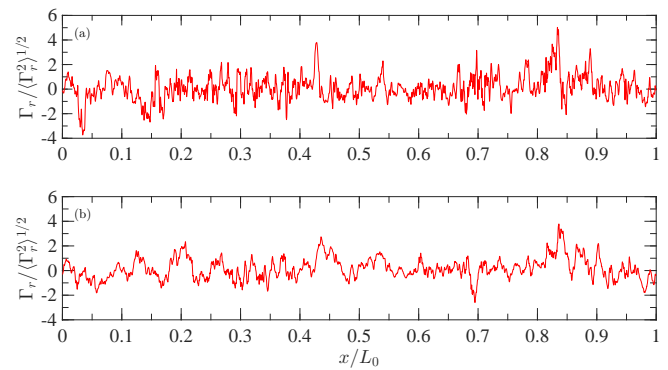}
\protect\caption{Typical circulation traces in the inertial range, for $\rel=1300$ (computational domain $8192^3$) along an arbitrary edge of the cube of length $L_0$. The linear size $r$ of the loops in the panels are, (a) $\nr=50$ and (b) $\nr=150$, corresponding to the lower and middle of the $4/5$-ths plateau in the normalized third-order structure function \cite{K41b}, respectively. 
}
\label{trace.fig}
\end{figure}

One other comment is useful. Circulation $\gr$ around a loop of side $r$ is calculated using the second equality in Eq.~\ref{def.eq}, which follows from the Stokes theorem, as the two-dimensional local average of vorticity using the algorithm given in Ref.~\cite{kp2014}. Statistical averages were taken over the whole $N^3$ simulation along the three
Cartesian directions. Statistics of $\gr$ were also calculated using the loop integration of Eq.~\ref{def.eq}, by means of cubic splines for improved accuracy, and excellent confirmation of the area integral results were obtained for $\nr > 5$.  


\section{Results}

\subsection{Area rules}
Figure \ref{trace.fig} shows circulation traces for inertial range separations, normalized by its standard deviation, along the length of the simulation domain of size $L_0$, at the Taylor scale Reynolds number, $\rel=1300$. Two different inertial separations $r$ corresponding to the lower end and the middle of the inertial range plateau in $-\dlllnr$, are shown here. These particular signals do not show frequent excursions to very high values (unlike highly intermittent phenomena).

\begin{figure}
\centering
\includegraphics[width=0.99\linewidth]{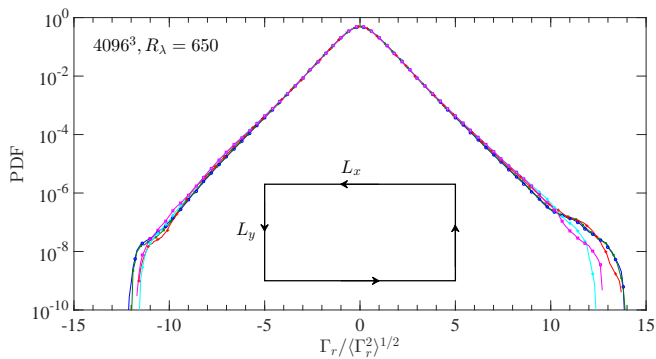}
\protect\caption{This figure verifies Migdal’s area rule. For five rectangles of different aspect ratios but the same area, all of which are contained in the inertial range, the circulation PDF are essentially invariant (the departures at the tails is mostly due to paucity of data). The different curves correspond to contour dimensions given by $L_x \times L_y$ (in units of $\eta$, the Kolmogorov length scale):
$(\colb{\circ})$ $30 \times 128$, 
$(\colg{\times})$ $32 \times 120$, 
$(\colr{+})$ $40 \times 96$, 
$(\colc{\ast})$ $64 \times 60$, 
$(\colm{\Box})$ $80 \times 48$. 
}
\label{shape.fig}
\end{figure}

\begin{figure}
\centering
\includegraphics[width=0.99\linewidth]{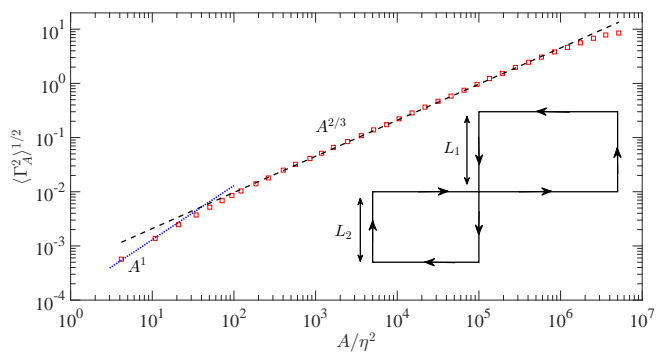}
\protect\caption{The circulation standard deviation $\la\Gamma_A^2 \ra^{1/2}$ $(\colr{\Box})$ as a function of the area $A$ of the figure-$8$ contour (shown in the inset), calculated on velocity data in a $4096^3$ grid at $\rel = 650$. The data follow the scalar area law, as evidenced by the $A^{2/3}$ scaling in the inertial range. The linear area dependence in the small-$A$ regime is shown, and the tendency to saturation for large areas is also apparent.
}
\label{loop8.fig}
\end{figure}

Figure \ref{shape.fig} shows the PDFs of $\Gamma_r$ for a number of rectangular loops of the same area but differing aspect ratios. The data collapse to a very good accuracy. Even though the result will have to be confirmed for loops of different shapes, this figure supports the expectation that only the area of the loop, not its shape, decides the PDF of circulation. It is thus sometimes convenient to use the symbol $\Gamma_A$ to describe the circulation around a loop of area $A$; henceforth, when we speak of $\Gamma_r$, it means circulation around a square of side $r$.

The inset to \ref{loop8.fig} shows a figure-$8$ loop with two different squares touching at a common vertex. If K41 is valid, it readily follows that the standard deviation of circulation $\langle \Gamma_A^2 \rangle^{1/2} \sim (L_1^2 + L_2^2)^{2/3} = A^{2/3}$. On the other hand, if one traverses along the loop in the direction of the arrows marked on the loop, the areas circumscribed by the two squares will have different signs. If the resulting vector area is the right quantity to use, the standard deviation $\langle \Gamma_A^2 \rangle^{1/2}$ for the figure-$8$ loop should scale as 
\beq
\label{vector.eq}
\langle \Gamma_A^2 \rangle^{1/2} \sim (L_1^2 - L_2^2)^{2/3} = (\Delta (L_1+L_2))^{2/3} 
\sim A^{1/3} \;,
\eeq
since, $\Delta = L_1 - L_2$ is a fixed constant. The main part of Fig.~\ref{loop8.fig} shows convincingly that the mean-square circulation varies as $A^{2/3}$, where $A$ is the scalar sum of the areas of the two loops, for most of the range. For small $A$, the variance is clearly linear in $A$ as expected from Taylor expansion, whereas it saturates for large $A$ because of many cancellations.

\subsection{Application of Kolmogorov's similarity argument}
\begin{figure}
\centering
\includegraphics[width=0.99\linewidth]{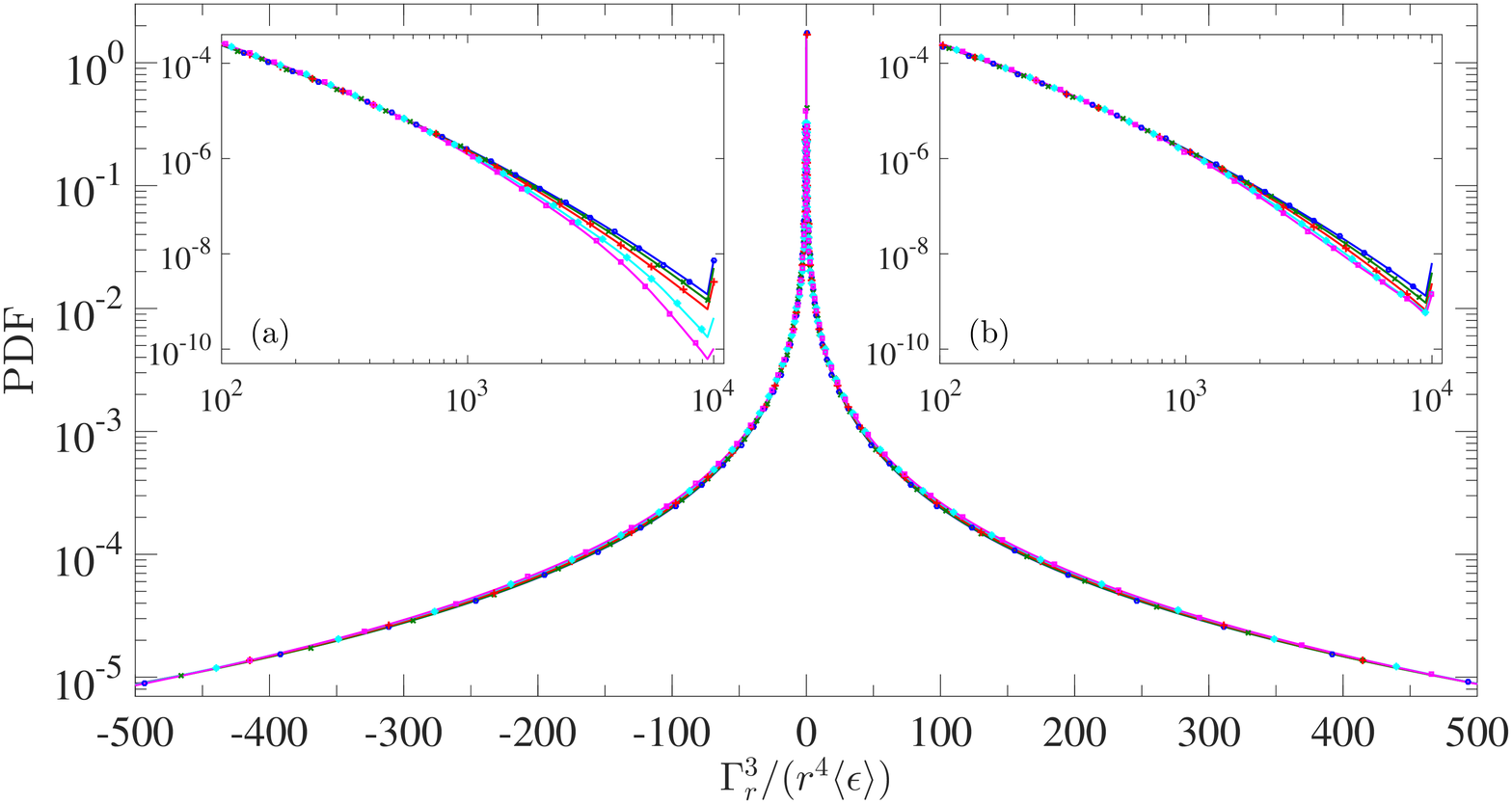}
\protect\caption{PDF of $\gcube$ at $\rel=1300$ at different inertial range separations, 
$\nr = 80$ $(\colb\circ)$,
$\nr = 120$ $(\colg\times)$,
$\nr = 171$ $(\colr{+})$,
$\nr = 246$ $(\colc\ast)$ and
$\nr = 300$ $(\colm\Box)$.
Insets (a) and (b) show portions of the PDF corresponding to negative and positive $\gr$, respectively, on log-log scales. The PDFs nearly collapse across the inertial range separations, with some differences towards the tails, as shown in the insets. The tails of each of the PDFs can be fitted by stretched exponentials with unequal stretching factors for the negative and positive tails. For the average curves, the negative part decays with a stretch factor of about 0.35, about twice as large as that for the positive ($\approx 0.17$). 
}
\label{pdfcube.fig}
\end{figure}

As already stated, straightforward application of K41 shows that $\Gamma_r \sim r^{4/3}$. Migdal \cite{migdal} argued that the PDF of $\gr$ decays as some power of ${\gr}^3/r^4$. We show the PDF of ${\gr^3}/{r^4}\epsm$ in Fig.~\ref{pdfcube.fig} for various inertial range separations. The PDFs show good collapse, with some deviations in the negative PDF tails; the two insets, highlighting the positive and negative tails of the PDF, are not power laws. They can be fitted nominally by unequal stretched exponentials, as described in the caption to Fig.~\ref{pdfcube.fig}. The mean and the mean square of this distribution yield the third-order and sixth-order moments of circulation. These and other moments are computed separately in Sec.~\ref{scexp.sec}.

The results so far are already interesting and tantalizing because they contain the suggestion that K41 may have currency for circulation. This would be quite unlike, say, for velocity increments or enstrophy density. We now explore this feature further by computing the scaling properties. 

\section{Scaling results}
\subsection{Exponents} \label{scexp.sec}
\begin{figure}
\centering
\includegraphics[width=0.99\linewidth]{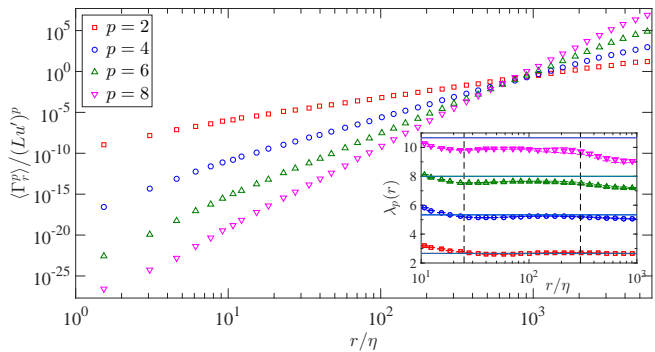}
\protect\caption{Normalized circulation moments for various orders $p$ as a function of the linear dimension of the loop, for $\rel = 1300$. Inset shows corresponding local slopes $\zp = d[\log \la \gr^p \ra]/d[\log r]$, where vertical lines demarcate the inertial range. Error bars (which are often subsumed by the symbol thickness) indicate $95\%$ confidence intervals. Horizontal straight lines drawn at $4p/3$ correspond to the respective K41 exponents.
}
\label{mom.fig}
\end{figure}
We first evaluate various even order moments of $\gr$ and show moments $\la \gr^p \ra$ for $p=2,4,6$ and $8$ for $\rel=1300$ as functions of $r$ in Fig.~\ref{mom.fig}. They all display proper power laws, $\la \gr^p \ra \sim r^{\lambda_p}$, in the inertial range, as reinforced by the constancy of the local slopes, shown in the inset of Fig.~\ref{mom.fig}. 

The odd moments of $\gr$ do not display equally clean power laws because they have negligible intensity via cancellation, which leads to poor convergence. Their scaling improves if one considers absolute values of circulation, $\la |\gr|^p \ra \sim r^\zpm$ (while, obviously, those of even orders remain unchanged). Absolute values enable us to define scaling exponents for fractional $p$ as well, up to (but not including) $p = -1$, for which the moments diverge \cite{chen05}. The practice of using absolute moments for odd orders is justified, at least {\it{a posteriori}}, as long as the exponents so obtained are monotonic when plotted together with even-order data. This monotonic behavior is not quite true for velocity structure functions even for very high Reynolds numbers typical of atmospheric flows (see Ref.~\cite{dhruva} at $\rel=10,340$) but seems to hold for circulation for which the absolute moment data do not zigzag with respect to even-order data. 

\begin{figure}
\centering
\includegraphics[width=0.99\linewidth]{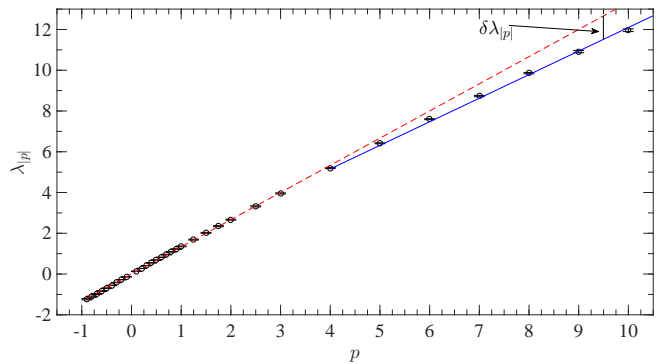}
\protect\caption{Scaling exponents as a function of moment order $p$; $R_\lambda = 1300$. The data can be fitted by two separate lines, one below order $3$ and one above it. The low order data can be fitted by the expression $4p/3$ (dashed line), which results from the application of K41. The high-order data can be fitted (solid line) by a fractal model (Eq.~\ref{monof.eq}) with $D \approx 2.5$. Least square fits are shown. Typical error bar shown for the tenth moment is subsumed by the symbol thickness. The difference between the K41 line and the measured exponents is indicated by $\delta\lambda_{|p|}$, which will be discussed in Sec.~\ref{discussion.sec}.
}
\label{circeven.fig}
\end{figure}

In Fig.~\ref{circeven.fig} we plot the scaling exponents $\zpm$ for circulation as a function of the power index $p$ for all orders at $\rel=1300$. The exponents seem to organize themselves into two straight lines, one below about 3 and the other above it. The low-order data can be fitted asymptotically by the K41 line, $\zpm = 4p/3$, but one needs a different line with a smaller slope for higher orders. The latter line (with the slope of 7/6 and an intercept of 1/2) can be expressed in terms of the $\beta$-model \cite{FSN78},
\beq
\label{monof.eq}
\zpm = \frac{4p}{3}+(3-D)(1-\frac{p}{3})\;,
\eeq
with $D \approx 2.5$ as the self-similarity dimension \cite{mandelbrot1974,FSN78}. 

We have so far considered results for the two highest Reynolds numbers of the dataset computed by us. It is instructive to examine how the results depend on the Reynolds number, and assess the asymptotic state. 

\subsection{Probability density functions}

\begin{figure}
\centering
\includegraphics[width=0.99\linewidth]{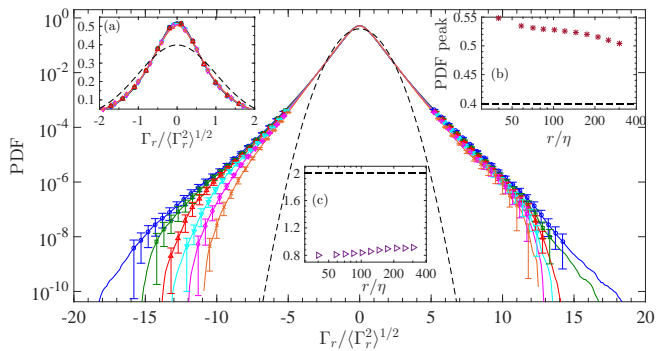}
\protect\caption{Probability density function of $\gr$ normalized by its standard deviation at $\rel=1300$, for inertial range separations
$\nr = 119$ $(\colb\circ)$,
$\nr = 143$ $(\colg\Box)$,
$\nr = 171$ $(\colr\triangle)$,
$\nr = 206$ $(\colc\triangledown)$,
$\nr = 246$ $(\colm\diamond)$ and
$\nr = 296$ $(\colo\times)$. Error bars shown for select large $|\gr|$ (for clarity)
 indicate $95\%$ confidence intervals. Dashed line is the standard Gaussian PDF. Insets: (a) expanded view of the PDF core to show deviations from Gaussianity, (b) PDF peaks {\it{vs}} $\nr$ in the inertial range, dashed line is the Gaussian peak and (c) stretching exponents of the positive PDF tails {\it{vs}} $\nr$.}
\label{pdf.fig}
\end{figure}

We have already seen the PDFs in two forms: The PDF for circulation around various shapes of loops within the inertial range in Fig.~\ref{shape.fig} and those for $\gcube$ in Fig.~\ref{pdfcube.fig}. We now plot the PDFs of $\gr$ for several values of $r$ in the inertial range, normalized by their own standard deviations. The PDFs collapse on each other for magnitudes below about 5 standard deviations (see inset (a)) and differ mostly in the tails. This shows that we do not have perfect self-similarity. The PDFs also depart strongly from the Gaussian distribution indicated by the dashed line, both at the tails and the core region (see inset (a)). The peaks of the PDFs gradually drop towards the Gaussian value of about 0.4 through the inertial range, as can be seen more explicitly in inset (b), but the tails are of the stretched exponential type $\sim exp(-\beta \gr^{\ar})$. The stretching exponent $\alpha(r)$ is plotted in inset (c) of Fig.~\ref{pdf.fig} as a function of the inertial range separation $r$; we have $\alpha(r)$ varying weakly with $r$, supporting the notion that circulation has weak intermittency \cite{vainshtein94}. We note that the variation of $\alpha(r)$ in the inertial range is gentler than that of the corresponding stretching exponents of velocity increments, which are known to rapidly change with scale \cite{krs92}. 

\subsection{Flatness}
\begin{figure}
\centering
\includegraphics[width=0.99\linewidth]{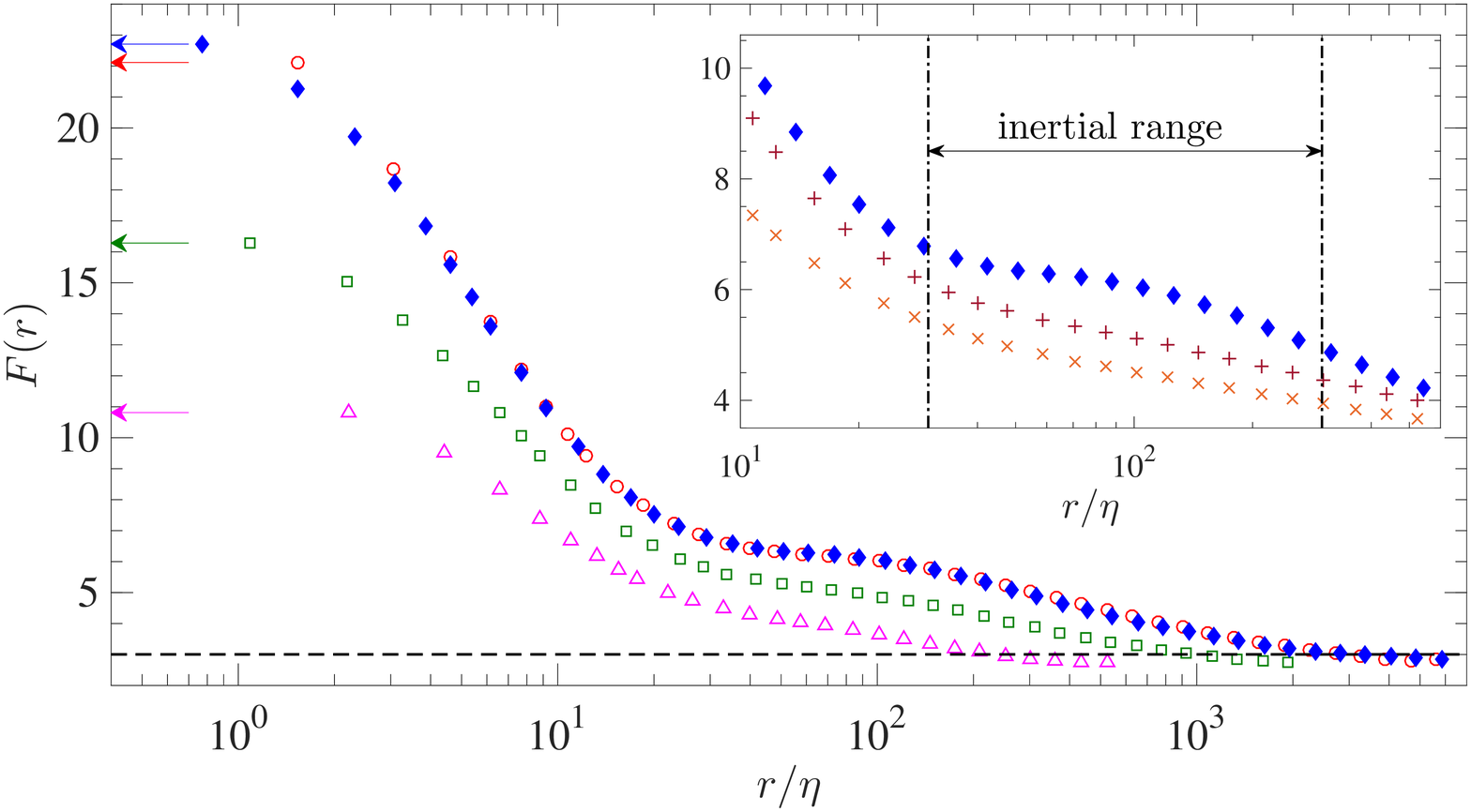}
\protect\caption{Circulation flatness $F(r)$ {\it{vs}}.~spatial separation $r$ from DNS at $\rel=240$, $512^3$ ($\colm{\triangle}$), $\rel=650$, $4096^3$ ($\colg{\Box}$) and $\rel=1300$, $8192^3$ ($\colr{\circ}$). The flatness from the DNS at $\rel=1300$ computed on a $16,384^3$ box is also shown ($\colb{\blacklozenge}$) and agrees well with $F(r)$ from the $8192^3$ DNS, for $r/\eta > 10$. Arrows on the ordinate show the limit $F(r \to 0) = \la\omega^4 \ra/\la \omega^2 \ra^2$, which is the flatness of the vorticity component normal to the circulation plane. At the largest scales, $F(r)$ is equal to the Gaussian flatness of $3$, shown by the dashed line. Inset compares the circulation flatness with those of the longitudinal ($\colo{\times}$) and transverse ($\colbr{+}$) velocity increments at $\rel=1300$ in the inertial range, shown by the dashed vertical lines. Unlike circulation with a tendency towards constancy in the inertial range, the velocity increments show gradual variations through the inertial range.
}
\label{flat.fig}
\end{figure}
\begin{figure}
\centering
\includegraphics[width=.8\linewidth]{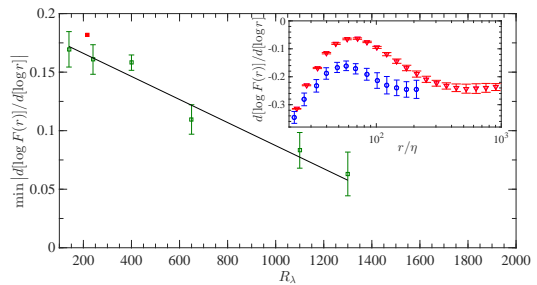}
\protect\caption{The peak magnitude of logarithmic local slope of circulation flatness in the inertial range, plotted against $\rel$. The solid line is a least-square fit. If extrapolated, the minimum of the logarithmic derivative will reach zero, corresponding to constant flatness in Fig.~\ref{flat.fig}, at $\rel \approx 1900$. Data of Ref.~\cite{CSKRS96} at $\rel=216$ ($\colr{\blacksquare}$), is shown for comparison. Inset shows the logarithmic local slopes of circulation flatness as a function of $\nr$, up to $r \sim O(L)$ for $\rel=1300$ ($\colm{\triangledown}$) and $\rel=240$ ($\colb{\circ}$). Error bars indicate $95\%$ confidence intervals. 
}
\label{minflat.fig}
\end{figure}
The circulation flatness is defined as
\beq
F(r) \equiv \frac{\la \gr ^4 \ra}{\la \gr^2 \ra^2} \;.
\eeq
Figure~\ref{flat.fig} shows that $F(r)$ varies through the inertial range $(\eta \ll r \ll L)$ but its evolution with Reynolds number is the more interesting point; the figure also compares $F(r)$ at three different Reynolds numbers. At $\rel = 240$, $F(r)$ increases smoothly from the Gaussian value of $3$, for $r \sim O(L)$, to the dissipation range limit of $\la \omega^4\ra/\la \omega^2 \ra^2$, where $\omega$ is the vorticity component normal to the plane of circulation. With increasing $\rel$, however, $F(r)$ appears to develop a plateau in the inertial range, indicating that it may be approaching a constant, independent of scale. The inset of Fig.~\ref{flat.fig} compares the circulation flatness to those of the longitudinal and transverse velocity increments at $\rel = 1300$. Here, the transverse increment $\dvr \equiv v(x+r)-v(x)$, where the separation distance $r$ is transverse to the velocity component $v$. Even at $\rel=1300$, the flatness factors of both $\dur$ and $\dvr$ smoothly grow with decreasing scale, showing that the velocity increments are highly intermittent, whereas $F(r)$ displays the tendency towards constancy, suggesting that $\gr$ at high Reynolds numbers is only weakly intermittent. It should be stressed that, at lower $\rel$, all flatness factors of $\gr$, $\dur$ and $\dvr$ increase rapidly with decreasing scale in the inertial range, as already shown in Refs.~\cite{CSKRS96,zhou08}.

Our point is that the flatness $F(r)$ has the potential to become a constant in the inertial range of $r$ as the Reynolds number increases further. If so, this will be the intermittency-free limit in which we expect the logarithmic local slope of flatness $d[\log F(r)]/d[\log r] \to 0$ in the inertial range. To quantify this approach to the intermittency free limit, we plot the logarithmic local slope of $F(r)$ as a function of spatial separation in the inset of Fig.~\ref{minflat.fig} for two different Reynolds numbers. For $\rel=1300$, the local slopes are closer to zero over a wider range of inertial scales than for $\rel=240$. The peaks of the flatness local slopes decrease linearly with $\rel$ as shown in Fig.~\ref{minflat.fig}, suggesting that it will be zero at $\rel \approx 1900$. This result should be regarded merely as an indication and will be discussed further in Sec.~\ref{asympstate.sec} below.
\section{Discussion} \label{discussion.sec}
Small-scale turbulence is known to be characterized by extreme events in time and space. Such intense events can be seen in local velocity increments, which exhibit intense spatial fluctuations, resulting in scaling exponents that vary nonlinearly with respect to the moment order. This is the phenomenon of intermittency that necessitates the superposition of infinitely many scale-invariant configurations to describe the exponents. Much effort has been expended in quantifying the nonlinear trend of the intermittency exponents of velocity moments, with varying degrees of success \cite{K62,menevkrs87,SL94,yakhot01}. As noted in Ref.~\cite{migdal95}, intermittency corrections from the velocity moments serve as an upper estimate, since the velocity field is infrared divergent. Here, using the largest simulations of isotropic turbulence to-date, we have shown that the structure of velocity circulation is much simpler at higher Reynolds numbers. In contrast to those of velocity increments, the circulation statistics become less intermittent with increasing Reynolds numbers, with the exponents having an approximately bifractal structure: at the level of low order moments, circulation is essentially a space-filling quantity whereas, for moments of order 3 and higher, it appears to have a self-similar dimension $D \approx 2.5$. 
  
These findings brighten the prospect of a simplified turbulence theory that can be used to describe small-scale turbulence without having to resort to multifractal models. We have inferred that circulation, though arising from highly convoluted vortex structures in space and time, is essentially space filling for low-order moments and can give rise to a bifractal scaling. It will be interesting to obtain circulation statistics in more realistic cases of anisotropic turbulence, to see if a statistical theory based on vortex filaments can shed new light on topics such as anisotropy effects on small-scale universality. 

Two questions appear worth discussing in some detail. (a) Since the data reveal that there is a Reynolds number dependence of the circulation properties, it is worth asking whether they have reached its asymptotic state even at the highest Reynolds number considered here, and, if not, make an educated guess on that state. This point is essentially an expansion of the tentative result deduced from Fig.~\ref{minflat.fig}. (b) Since circulation is very closely related to vorticity, it is natural to ask why the simplicity that is apparent in circulation does not translate to enstrophy density, which is known to be a strongly multifractal quantity \cite{MSF90}. We will discuss these questions in that same order.
\begin{figure}
\centering
\includegraphics[width=.99\linewidth]{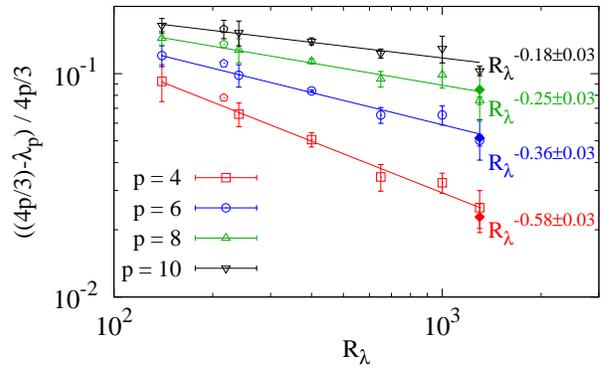}
\protect\caption{The relative difference of the circulation scaling exponent $\zp$ from the K41 exponent, $4p/3$, as a function of the Taylor-scale Reynolds numbers $\rel$, for orders $p=4,6,8$ and $10$. Data from Ref.~\cite{CSKRS96} $(\pentagon)$ for orders $p = 4,6,8$ and $10$, at $\rel=216$ and the
highly resolved DNS for orders $p=4,6$ and $8$, at $\rel=1300$,$16384^3$ (${\blacklozenge}$), are provided for comparison. Exponential least square fits through the data are shown by the solid lines; also shown are error bars.
}
\label{expdevk41.fig}
\end{figure}
\subsection{The asymptotic state} \label{asympstate.sec}
Figure \ref{expdevk41.fig} shows the behavior of $\delta_{\lambda_p}$, the difference between the measured scaling exponents and the corresponding Kolmogorov values, for moment orders $4$, $6$, $8$ and $10$, as functions of the Reynolds number. For each order, this difference seems to approach K41 roughly as a power law. The rates of approach vary inversely with the order of the moment. We cannot speculate about the behaviors at infinitely large Reynolds numbers, but may expect that this behavior will persist up to some higher Reynolds number. The fact that the approach to the K41 values is slower for the high-order moments suggests that, in principle, the scaling will remain a bifractal for all finite Reynolds numbers, with the ``phase transition" point moving to a higher $p$ with increasing Reynolds numbers. In practice, however, moments may fall on the K41 line for sufficiently high values of $p$ that the bifractal behavior may essentially yield place to a space-filling monofractal. 

\subsection{Circulation, enstrophy and velocity increments}
Since circulation is very closely related to vorticity, it is natural to ask why the simplicity that is apparent in circulation does not translate to locally averaged enstrophy density, which is known to be a strongly multifractal quantity \cite{MSF90,YSP18}. The corresponding question is relevant also with respect to velocity increments. We address this issue here briefly.

The area integral formula for circulation from Eq.~\ref{def.eq} reads as
\beq
\circr = \int \omega dA \;,
\eeq
where $\omega$ is the vorticity component perpendicular to the loop with area, $A = r^2$, along an arbitrary direction.  Taking the square of above equation and invoking Schwarz's inequality, we get
\beqa
\circr^2  &=& \Big [\int \omega dA \Big ]^2 \;, \\
          &\le & r^2 \int \omega^2 dA   \;.
\eeqa
Taking averages on both sides of the above inequality we get
\beq
\label{cineq.eq}
\la \circr^2 \ra \le r^2 \Big\la \int \omega^2 dA \Big\ra \;.
\eeq
Now, define $\Omega = \omega_i \omega_i$ and use statistical isotropy. Then, for any general power, $p \ge 1$,
\beq
\label{omgineq.eq}
\Big \la \Big ( \int \omega^2 dA \Big )^{p} \Big \ra \le \frac{1}{3} \Big\la \Big ( \int \Omega dA \Big )^{p} \Big \ra, \
\eeq
with equality valid for $p=1$.
Substituting Eq.~\ref{omgineq.eq} in Eq.~\ref{cineq.eq} for
$p=1$, we obtain
\beq
\label{enst.eq}
\la \circr^2 \ra \le \frac{1}{3} r^2 \Big\la \int \Omega dA \Big\ra \;\;.
\eeq 
We can define the local 2D average of enstrophy as
\beq
\Omega_r \equiv \frac{1}{r^2} \int \Omega dA \;,
\eeq 
and rewrite Eq.~\ref{enst.eq} as
\beq
\label{ord2.eq}
\la \circr^2 \ra  \le \frac{1}{3} r^4 \la \Omega_r \ra \;. \\
\eeq
Generalizing the above inequality to any even power $2p$, we get
\beq
\label{relomg.eq}
\la \circr^{2p} \ra  \le \frac{1}{3} {r}^{4p} \la \Omega_r^{p} \ra \;. \\
\eeq
This inequality shows that the moments of locally averaged enstrophy are quite likely to be larger than those of circulation, and it is this that states that the multifractal character of enstrophy density need not necessarily carry over to circulation.

With respect to moments of the velocity increments, we can use the second hypothesis of Kolmogorov
\citep{K62} which can be expressed as
\beq
\Delta_r u = V(r\epsilon_r)^{1/3} \equiv V'(r\nu \Omega_r)^{1/3},
\eeq
where $V$ and $V'$ are universal variables with the constraint that,
 $\la V^3 \ra = \la V'^3 \ra = -4/5$, and $\nu$ (as before) is the kinematic viscosity.
Then, Eq.~\ref{relomg.eq} becomes
\beq
\label{k62.eq}
\la \circr^{2p} \ra  \le \frac{1}{3}{r}^{3p} \frac{\la (\Delta_r u)^{3p} \ra}{\nu^p \la V'^{3p} \ra} \;. \\
\eeq

Under homogeneity, the first moments of dissipation, enstrophy and their local averages are related as $\nu \la \Omega_r \ra =  \la \epsilon_r \ra = \meandiss$, using which, inequality (14) (corresponding to $p=1$) can be written as
\beq
\label{cbound.eq}
\psi(r) \equiv \frac{\la \circr^{2} \ra \nu}{r^4 \meandiss}  \le \frac{1}{3} \;. \\
\eeq
Using $p=1$ and the inertial range result $\la (\Delta_r u)^3 \ra = (-4/5)r\meandiss$, we end up with the same inequality as Eq. \ref{cbound.eq}. Inequality \ref{cbound.eq} is more general, since it shows that $\psi(r)$ is bounded for all $r$.
\begin{figure}
\centering
\includegraphics[width=.8\linewidth]{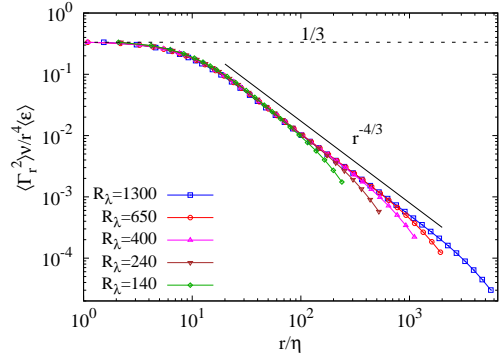}
\protect\caption{The function $\psi(r)$ (left hand side of inequality \ref{cbound.eq}) as a function of normalized spatial separation $r/\eta$ for different Reynolds numbers. The curves are seen to collapse, for $r \ll L$ ($L$ is the integral scale) at the different Reynolds numbers shown, reasonably well. The limit $\psi(r\to 0) = 1/3$ is marked by the horizontal dashed line. Solid black line indicates the corresponding K41 slope.
}
\label{cbound.fig}
\end{figure} 
As a check, we note that in the dissipative limit $r \to 0$, we have
\beq
\lim_{r \to 0} \psi(r) = 
\lim_{r \to 0} \frac{\la \circr^{2} \ra\nu}{r^4\meandiss} = \frac{\la \omega^2 \ra \nu}{\meandiss} 
= \frac{1}{3}\frac{\la \Omega \ra \nu}{\meandiss} = \frac{1}{3} \;,
\eeq
where the relation $\la \Omega \ra = 3\la\omega^2\ra$ uses isotropy.
 
It is interesting to consider the double limit of $\psi(r)$, namely
\beq
\lim_{\nu \to 0} \lim_{r \to 0} \frac{\la \circr^{2} \ra}{r^4} \left [\frac{\nu}{\meandiss} \right ] \;.
\eeq
Since the mean dissipation is known not to depend on viscosity \cite{KRS84,KRS98,pearson2002,kaneda03}, the term within square brackets that is $r$-independent, tends to $0$ in the asymptotic limit. This means that in the small-$r$ limit we have, $\la \circr^{2} \ra/r^4 = \la \omega^2 \ra \to \infty$ for high Reynolds numbers. It follows that vorticity can attain very high moment values in turbulent flows even when those of circulation are bounded. This observation does not necessarily imply that circulation (around finite contours) is always bounded, but it does suggest that it is a more tractable quantity to work with, as compared to vorticity.
  
Figure \ref{cbound.fig} shows that the DNS data are quite consistent with the dissipative limit. Furthermore, $\psi(r)$ at different Reynolds numbers shows good collapse for $r \ll L$, where $L$ is the integral scale.

These arguments show that the simplicity in the structure of circulation does not contradict the known multifractal properties of velocity increments or enstrophy density. We thus believe that we have covered new ground here.
\section{Acknowledgments}
This work is partially supported by the National Science Foundation (NSF), via Grants No.~ACI-$1640771$  and No.~ACI-$1036170$ at the Georgia Institute of Technology. The computations were performed using supercomputing resources provided through the XSEDE consortium (which is funded by NSF) at the Texas Advanced Computing Center at the University of Texas (Austin), and the Blue Waters Project at the National Center for Supercomputing Applications at the University of Illinois (Urbana-Champaign). 
We thank Dr.~Xiaomeng Zhai for his help in evaluating circulation via spline fits, and Drs.~A.~A.~Migdal and V.~Yakhot for valuable discussions over time.  
\bibliography{zebib}

\begin{thebibliography}{38}%
\makeatletter
\providecommand \@ifxundefined [1]{%
 \@ifx{#1\undefined}
}%
\providecommand \@ifnum [1]{%
 \ifnum #1\expandafter \@firstoftwo
 \else \expandafter \@secondoftwo
 \fi
}%
\providecommand \@ifx [1]{%
 \ifx #1\expandafter \@firstoftwo
 \else \expandafter \@secondoftwo
 \fi
}%
\providecommand \natexlab [1]{#1}%
\providecommand \enquote  [1]{``#1''}%
\providecommand \bibnamefont  [1]{#1}%
\providecommand \bibfnamefont [1]{#1}%
\providecommand \citenamefont [1]{#1}%
\providecommand \href@noop [0]{\@secondoftwo}%
\providecommand \href [0]{\begingroup \@sanitize@url \@href}%
\providecommand \@href[1]{\@@startlink{#1}\@@href}%
\providecommand \@@href[1]{\endgroup#1\@@endlink}%
\providecommand \@sanitize@url [0]{\catcode `\\12\catcode `\$12\catcode
  `\&12\catcode `\#12\catcode `\^12\catcode `\_12\catcode `\%12\relax}%
\providecommand \@@startlink[1]{}%
\providecommand \@@endlink[0]{}%
\providecommand \url  [0]{\begingroup\@sanitize@url \@url }%
\providecommand \@url [1]{\endgroup\@href {#1}{\urlprefix }}%
\providecommand \urlprefix  [0]{URL }%
\providecommand \Eprint [0]{\href }%
\providecommand \doibase [0]{http://dx.doi.org/}%
\providecommand \selectlanguage [0]{\@gobble}%
\providecommand \bibinfo  [0]{\@secondoftwo}%
\providecommand \bibfield  [0]{\@secondoftwo}%
\providecommand \translation [1]{[#1]}%
\providecommand \BibitemOpen [0]{}%
\providecommand \bibitemStop [0]{}%
\providecommand \bibitemNoStop [0]{.\EOS\space}%
\providecommand \EOS [0]{\spacefactor3000\relax}%
\providecommand \BibitemShut  [1]{\csname bibitem#1\endcsname}%
\let\auto@bib@innerbib\@empty
\bibitem [{\citenamefont {da~Vinci}(2010)}]{vinci}%
  \BibitemOpen
  \bibfield  {author} {\bibinfo {author} {\bibfnamefont {L.}~\bibnamefont
  {da~Vinci}},\ }\href@noop {} {\emph {\bibinfo {title} {{The notebooks of
  Leonardo da Vinci}}}},\ edited by\ \bibinfo {editor} {\bibfnamefont {J.~P.}\
  \bibnamefont {Richter}}\ (\bibinfo  {publisher} {Pac PS},\ \bibinfo {year}
  {2010})\BibitemShut {NoStop}%
\bibitem [{\citenamefont {Kaneda}\ \emph {et~al.}(2003)\citenamefont {Kaneda},
  \citenamefont {Ishihara}, \citenamefont {Yokokawa}, \citenamefont {Itakura},\
  and\ \citenamefont {Uno}}]{kaneda03}%
  \BibitemOpen
  \bibfield  {author} {\bibinfo {author} {\bibfnamefont {Y.}~\bibnamefont
  {Kaneda}}, \bibinfo {author} {\bibfnamefont {T.}~\bibnamefont {Ishihara}},
  \bibinfo {author} {\bibfnamefont {M.}~\bibnamefont {Yokokawa}}, \bibinfo
  {author} {\bibfnamefont {K.}~\bibnamefont {Itakura}}, \ and\ \bibinfo
  {author} {\bibfnamefont {A.}~\bibnamefont {Uno}},\ }\href@noop {} {\bibfield
  {journal} {\bibinfo  {journal} {Phys. Fluids}\ }\textbf {\bibinfo {volume}
  {15}},\ \bibinfo {pages} {L21} (\bibinfo {year} {2003})}\BibitemShut
  {NoStop}%
\bibitem [{\citenamefont {Kolmogorov}(941a)}]{K41a}%
  \BibitemOpen
  \bibfield  {author} {\bibinfo {author} {\bibfnamefont {A.~N.}\ \bibnamefont
  {Kolmogorov}},\ }\href@noop {} {\bibfield  {journal} {\bibinfo  {journal}
  {Dokl. Akad. Nauk. SSSR}\ }\textbf {\bibinfo {volume} {30}},\ \bibinfo
  {pages} {299} (\bibinfo {year} {1941a})}\BibitemShut {NoStop}%
\bibitem [{\citenamefont {Monin}\ and\ \citenamefont {Yaglom}(1975)}]{MY.II}%
  \BibitemOpen
  \bibfield  {author} {\bibinfo {author} {\bibfnamefont {A.~S.}\ \bibnamefont
  {Monin}}\ and\ \bibinfo {author} {\bibfnamefont {A.~M.}\ \bibnamefont
  {Yaglom}},\ }\href@noop {} {\emph {\bibinfo {title} {Statistical Fluid
  Mechanics}}},\ Vol.~\bibinfo {volume} {2}\ (\bibinfo  {publisher} {MIT
  Press},\ \bibinfo {year} {1975})\BibitemShut {NoStop}%
\bibitem [{\citenamefont {Benzi}\ \emph {et~al.}(1984)\citenamefont {Benzi},
  \citenamefont {Paladin}, \citenamefont {Parisi},\ and\ \citenamefont
  {Vulpiani}}]{Benzi1984}%
  \BibitemOpen
  \bibfield  {author} {\bibinfo {author} {\bibfnamefont {R.}~\bibnamefont
  {Benzi}}, \bibinfo {author} {\bibfnamefont {G.}~\bibnamefont {Paladin}},
  \bibinfo {author} {\bibfnamefont {G.}~\bibnamefont {Parisi}}, \ and\ \bibinfo
  {author} {\bibfnamefont {A.}~\bibnamefont {Vulpiani}},\ }\href@noop {}
  {\bibfield  {journal} {\bibinfo  {journal} {J. Phys. A: Math. Gen.}\ }\textbf
  {\bibinfo {volume} {17}},\ \bibinfo {pages} {3521} (\bibinfo {year}
  {1984})}\BibitemShut {NoStop}%
\bibitem [{\citenamefont {Frisch}(1995)}]{Fri95}%
  \BibitemOpen
  \bibfield  {author} {\bibinfo {author} {\bibfnamefont {U.}~\bibnamefont
  {Frisch}},\ }\href@noop {} {\emph {\bibinfo {title} {Turbulence}}}\ (\bibinfo
   {publisher} {Cambridge University Press},\ \bibinfo {year}
  {1995})\BibitemShut {NoStop}%
\bibitem [{\citenamefont {Eyink}(1995)}]{Eyink1995}%
  \BibitemOpen
  \bibfield  {author} {\bibinfo {author} {\bibfnamefont {G.~L.}\ \bibnamefont
  {Eyink}},\ }\href@noop {} {\bibfield  {journal} {\bibinfo  {journal} {J.
  Stat. Phys.}\ }\textbf {\bibinfo {volume} {78}},\ \bibinfo {pages} {353}
  (\bibinfo {year} {1995})}\BibitemShut {NoStop}%
\bibitem [{\citenamefont {Sreenivasan}\ and\ \citenamefont
  {Antonia}(1997)}]{SA97}%
  \BibitemOpen
  \bibfield  {author} {\bibinfo {author} {\bibfnamefont {K.~R.}\ \bibnamefont
  {Sreenivasan}}\ and\ \bibinfo {author} {\bibfnamefont {R.~A.}\ \bibnamefont
  {Antonia}},\ }\href@noop {} {\bibfield  {journal} {\bibinfo  {journal} {Annu.
  Rev. Fluid Mech.}\ }\textbf {\bibinfo {volume} {29}},\ \bibinfo {pages} {435}
  (\bibinfo {year} {1997})}\BibitemShut {NoStop}%
\bibitem [{\citenamefont {Migdal}(1994)}]{migdal}%
  \BibitemOpen
  \bibfield  {author} {\bibinfo {author} {\bibfnamefont {A.~A.}\ \bibnamefont
  {Migdal}},\ }\href@noop {} {\bibfield  {journal} {\bibinfo  {journal} {Int.
  J. Mod. Phys. A}\ }\textbf {\bibinfo {volume} {9}},\ \bibinfo {pages} {1197}
  (\bibinfo {year} {1994})}\BibitemShut {NoStop}%
\bibitem [{\citenamefont {Umeki}(1993)}]{umeki}%
  \BibitemOpen
  \bibfield  {author} {\bibinfo {author} {\bibfnamefont {M.}~\bibnamefont
  {Umeki}},\ }\href@noop {} {\bibfield  {journal} {\bibinfo  {journal} {J.
  Phys. Soc. Jpn.}\ }\textbf {\bibinfo {volume} {69}},\ \bibinfo {pages} {3788}
  (\bibinfo {year} {1993})}\BibitemShut {NoStop}%
\bibitem [{\citenamefont {Sreenivasan}\ \emph {et~al.}(1995)\citenamefont
  {Sreenivasan}, \citenamefont {Juneja},\ and\ \citenamefont {Suri}}]{KRSJS95}%
  \BibitemOpen
  \bibfield  {author} {\bibinfo {author} {\bibfnamefont {K.~R.}\ \bibnamefont
  {Sreenivasan}}, \bibinfo {author} {\bibfnamefont {A.}~\bibnamefont {Juneja}},
  \ and\ \bibinfo {author} {\bibfnamefont {A.~K.}\ \bibnamefont {Suri}},\
  }\href@noop {} {\bibfield  {journal} {\bibinfo  {journal} {Phys. Rev. Lett.}\
  }\textbf {\bibinfo {volume} {75}},\ \bibinfo {pages} {433} (\bibinfo {year}
  {1995})}\BibitemShut {NoStop}%
\bibitem [{\citenamefont {Cao}\ \emph {et~al.}(1996)\citenamefont {Cao},
  \citenamefont {Chen},\ and\ \citenamefont {Sreenivasan}}]{CSKRS96}%
  \BibitemOpen
  \bibfield  {author} {\bibinfo {author} {\bibfnamefont {N.}~\bibnamefont
  {Cao}}, \bibinfo {author} {\bibfnamefont {S.}~\bibnamefont {Chen}}, \ and\
  \bibinfo {author} {\bibfnamefont {K.~R.}\ \bibnamefont {Sreenivasan}},\
  }\href@noop {} {\bibfield  {journal} {\bibinfo  {journal} {Phys. Rev. Lett.}\
  }\textbf {\bibinfo {volume} {76}},\ \bibinfo {pages} {616} (\bibinfo {year}
  {1996})}\BibitemShut {NoStop}%
\bibitem [{\citenamefont {Benzi}\ \emph {et~al.}(1997)\citenamefont {Benzi},
  \citenamefont {Biferale}, \citenamefont {Struglia},\ and\ \citenamefont
  {Tripiccione}}]{benzi97}%
  \BibitemOpen
  \bibfield  {author} {\bibinfo {author} {\bibfnamefont {R.}~\bibnamefont
  {Benzi}}, \bibinfo {author} {\bibfnamefont {L.}~\bibnamefont {Biferale}},
  \bibinfo {author} {\bibfnamefont {M.~V.}\ \bibnamefont {Struglia}}, \ and\
  \bibinfo {author} {\bibfnamefont {R.}~\bibnamefont {Tripiccione}},\
  }\href@noop {} {\bibfield  {journal} {\bibinfo  {journal} {Phys. Rev. E}\
  }\textbf {\bibinfo {volume} {55}},\ \bibinfo {pages} {3739} (\bibinfo {year}
  {1997})}\BibitemShut {NoStop}%
\bibitem [{\citenamefont {Zhou}\ \emph {et~al.}(2008)\citenamefont {Zhou},
  \citenamefont {Sun},\ and\ \citenamefont {Xia}}]{zhou08}%
  \BibitemOpen
  \bibfield  {author} {\bibinfo {author} {\bibfnamefont {Q.}~\bibnamefont
  {Zhou}}, \bibinfo {author} {\bibfnamefont {C.}~\bibnamefont {Sun}}, \ and\
  \bibinfo {author} {\bibfnamefont {K.~Q.}\ \bibnamefont {Xia}},\ }\href@noop
  {} {\bibfield  {journal} {\bibinfo  {journal} {J. Fluid Mech.}\ }\textbf
  {\bibinfo {volume} {598}},\ \bibinfo {pages} {361–372} (\bibinfo {year}
  {2008})}\BibitemShut {NoStop}%
\bibitem [{\citenamefont {Yeung}\ \emph {et~al.}(2015)\citenamefont {Yeung},
  \citenamefont {Zhai},\ and\ \citenamefont {Sreenivasan}}]{pkpnas}%
  \BibitemOpen
  \bibfield  {author} {\bibinfo {author} {\bibfnamefont {P.~K.}\ \bibnamefont
  {Yeung}}, \bibinfo {author} {\bibfnamefont {X.~M.}\ \bibnamefont {Zhai}}, \
  and\ \bibinfo {author} {\bibfnamefont {K.~R.}\ \bibnamefont {Sreenivasan}},\
  }\href@noop {} {\bibfield  {journal} {\bibinfo  {journal} {Proc. Nat. Acad.
  Sci.}\ }\textbf {\bibinfo {volume} {112}},\ \bibinfo {pages} {12633}
  (\bibinfo {year} {2015})}\BibitemShut {NoStop}%
\bibitem [{\citenamefont {Yeung}\ \emph {et~al.}(2018)\citenamefont {Yeung},
  \citenamefont {Sreenivasan},\ and\ \citenamefont {Pope}}]{YSP18}%
  \BibitemOpen
  \bibfield  {author} {\bibinfo {author} {\bibfnamefont {P.~K.}\ \bibnamefont
  {Yeung}}, \bibinfo {author} {\bibfnamefont {K.~R.}\ \bibnamefont
  {Sreenivasan}}, \ and\ \bibinfo {author} {\bibfnamefont {S.~B.}\ \bibnamefont
  {Pope}},\ }\href@noop {} {\ \textbf {\bibinfo {volume} {3}},\ \bibinfo
  {pages} {064603} (\bibinfo {year} {2018})}\BibitemShut {NoStop}%
\bibitem [{\citenamefont {Kolmogorov}(941b)}]{K41b}%
  \BibitemOpen
  \bibfield  {author} {\bibinfo {author} {\bibfnamefont {A.~N.}\ \bibnamefont
  {Kolmogorov}},\ }\href@noop {} {\bibfield  {journal} {\bibinfo  {journal}
  {Dokl. Akad. Nauk. SSSR}\ }\textbf {\bibinfo {volume} {434}},\ \bibinfo
  {pages} {16} (\bibinfo {year} {1941b})}\BibitemShut {NoStop}%
\bibitem [{\citenamefont {Iyer}\ \emph {et~al.}(2017)\citenamefont {Iyer},
  \citenamefont {Sreenivasan},\ and\ \citenamefont {Yeung}}]{KI16}%
  \BibitemOpen
  \bibfield  {author} {\bibinfo {author} {\bibfnamefont {K.~P.}\ \bibnamefont
  {Iyer}}, \bibinfo {author} {\bibfnamefont {K.~R.}\ \bibnamefont
  {Sreenivasan}}, \ and\ \bibinfo {author} {\bibfnamefont {P.~K.}\ \bibnamefont
  {Yeung}},\ }\href@noop {} {\bibfield  {journal} {\bibinfo  {journal} {Phys.
  Rev. E}\ }\textbf {\bibinfo {volume} {95}},\ \bibinfo {pages} {021101}
  (\bibinfo {year} {2017})}\BibitemShut {NoStop}%
\bibitem [{\citenamefont {Eswaran}\ and\ \citenamefont {Pope}(1988)}]{EP88}%
  \BibitemOpen
  \bibfield  {author} {\bibinfo {author} {\bibfnamefont {V.}~\bibnamefont
  {Eswaran}}\ and\ \bibinfo {author} {\bibfnamefont {S.~B.}\ \bibnamefont
  {Pope}},\ }\href@noop {} {\bibfield  {journal} {\bibinfo  {journal} {Comput.
  Fluids}\ }\textbf {\bibinfo {volume} {16}},\ \bibinfo {pages} {257} (\bibinfo
  {year} {1988})}\BibitemShut {NoStop}%
\bibitem [{\citenamefont {Donzis}\ and\ \citenamefont {Yeung}(2010)}]{DY10}%
  \BibitemOpen
  \bibfield  {author} {\bibinfo {author} {\bibfnamefont {D.~A.}\ \bibnamefont
  {Donzis}}\ and\ \bibinfo {author} {\bibfnamefont {P.~K.}\ \bibnamefont
  {Yeung}},\ }\href@noop {} {\bibfield  {journal} {\bibinfo  {journal} {Physica
  D}\ }\textbf {\bibinfo {volume} {{239}}},\ \bibinfo {pages} {1278} (\bibinfo
  {year} {2010})}\BibitemShut {NoStop}%
\bibitem [{\citenamefont {Rogallo}(1981)}]{rogallo}%
  \BibitemOpen
  \bibfield  {author} {\bibinfo {author} {\bibfnamefont {R.~S.}\ \bibnamefont
  {Rogallo}},\ }\href@noop {} {\bibfield  {journal} {\bibinfo  {journal} {NASA
  Tech.~{} Memo}\ } (\bibinfo {year} {1981})}\BibitemShut {NoStop}%
\bibitem [{\citenamefont {Ishihara}\ \emph {et~al.}(2009)\citenamefont
  {Ishihara}, \citenamefont {Gotoh},\ and\ \citenamefont {Kaneda}}]{IGK2009}%
  \BibitemOpen
  \bibfield  {author} {\bibinfo {author} {\bibfnamefont {T.}~\bibnamefont
  {Ishihara}}, \bibinfo {author} {\bibfnamefont {T.}~\bibnamefont {Gotoh}}, \
  and\ \bibinfo {author} {\bibfnamefont {Y.}~\bibnamefont {Kaneda}},\
  }\href@noop {} {\bibfield  {journal} {\bibinfo  {journal} {Annu. Rev. Fluid
  Mech.}\ }\textbf {\bibinfo {volume} {41}},\ \bibinfo {pages} {165} (\bibinfo
  {year} {2009})}\BibitemShut {NoStop}%
\bibitem [{\citenamefont {Iyer}(2014)}]{kp2014}%
  \BibitemOpen
  \bibfield  {author} {\bibinfo {author} {\bibfnamefont {K.~P.}\ \bibnamefont
  {Iyer}},\ }\emph {\bibinfo {title} {Studies of turbulence structure and
  turbulent mixing using Petascale computing}},\ \href@noop {} {\bibinfo {type}
  {{Ph.D.} thesis}},\ \bibinfo  {school} {Georgia Institute of Technology}
  (\bibinfo {year} {2014})\BibitemShut {NoStop}%
\bibitem [{\citenamefont {Chen}\ \emph {et~al.}(2005)\citenamefont {Chen},
  \citenamefont {Dhruva}, \citenamefont {Kurien}, \citenamefont {Sreenivasan},\
  and\ \citenamefont {Taylor}}]{chen05}%
  \BibitemOpen
  \bibfield  {author} {\bibinfo {author} {\bibfnamefont {S.~Y.}\ \bibnamefont
  {Chen}}, \bibinfo {author} {\bibfnamefont {B.}~\bibnamefont {Dhruva}},
  \bibinfo {author} {\bibfnamefont {S.}~\bibnamefont {Kurien}}, \bibinfo
  {author} {\bibfnamefont {K.~R.}\ \bibnamefont {Sreenivasan}}, \ and\ \bibinfo
  {author} {\bibfnamefont {M.~A.}\ \bibnamefont {Taylor}},\ }\href@noop {}
  {\bibfield  {journal} {\bibinfo  {journal} {J. Fluid Mech.}\ }\textbf
  {\bibinfo {volume} {533}},\ \bibinfo {pages} {183} (\bibinfo {year}
  {2005})}\BibitemShut {NoStop}%
\bibitem [{\citenamefont {Sreenivasan}\ and\ \citenamefont
  {Dhruva}(1998)}]{dhruva}%
  \BibitemOpen
  \bibfield  {author} {\bibinfo {author} {\bibfnamefont {K.~R.}\ \bibnamefont
  {Sreenivasan}}\ and\ \bibinfo {author} {\bibfnamefont {B.}~\bibnamefont
  {Dhruva}},\ }\href@noop {} {\bibfield  {journal} {\bibinfo  {journal} {Prog.
  Theor. Phys. Suppl.}\ }\textbf {\bibinfo {volume} {130}},\ \bibinfo {pages}
  {103} (\bibinfo {year} {1998})}\BibitemShut {NoStop}%
\bibitem [{\citenamefont {Frisch}\ \emph {et~al.}(1978)\citenamefont {Frisch},
  \citenamefont {Sulem},\ and\ \citenamefont {Nelkin}}]{FSN78}%
  \BibitemOpen
  \bibfield  {author} {\bibinfo {author} {\bibfnamefont {U.}~\bibnamefont
  {Frisch}}, \bibinfo {author} {\bibfnamefont {P.~L.}\ \bibnamefont {Sulem}}, \
  and\ \bibinfo {author} {\bibfnamefont {M.}~\bibnamefont {Nelkin}},\
  }\href@noop {} {\bibfield  {journal} {\bibinfo  {journal} {J. Fluid Mech.}\
  }\textbf {\bibinfo {volume} {87}},\ \bibinfo {pages} {719–736} (\bibinfo
  {year} {1978})}\BibitemShut {NoStop}%
\bibitem [{\citenamefont {Mandelbrot}(1974)}]{mandelbrot1974}%
  \BibitemOpen
  \bibfield  {author} {\bibinfo {author} {\bibfnamefont {B.~B.}\ \bibnamefont
  {Mandelbrot}},\ }\href@noop {} {\bibfield  {journal} {\bibinfo  {journal} {J.
  Fluid Mech.}\ }\textbf {\bibinfo {volume} {62}},\ \bibinfo {pages}
  {331–358} (\bibinfo {year} {1974})}\BibitemShut {NoStop}%
\bibitem [{\citenamefont {Vainshtein}\ \emph {et~al.}(1994)\citenamefont
  {Vainshtein}, \citenamefont {Sreenivasan}, \citenamefont {Pierrehumbert},
  \citenamefont {Kashyap},\ and\ \citenamefont {Juneja}}]{vainshtein94}%
  \BibitemOpen
  \bibfield  {author} {\bibinfo {author} {\bibfnamefont {S.~I.}\ \bibnamefont
  {Vainshtein}}, \bibinfo {author} {\bibfnamefont {K.~R.}\ \bibnamefont
  {Sreenivasan}}, \bibinfo {author} {\bibfnamefont {R.~T.}\ \bibnamefont
  {Pierrehumbert}}, \bibinfo {author} {\bibfnamefont {V.}~\bibnamefont
  {Kashyap}}, \ and\ \bibinfo {author} {\bibfnamefont {A.}~\bibnamefont
  {Juneja}},\ }\href@noop {} {\bibfield  {journal} {\bibinfo  {journal} {Phys.
  Rev. E}\ }\textbf {\bibinfo {volume} {50}},\ \bibinfo {pages} {1823}
  (\bibinfo {year} {1994})}\BibitemShut {NoStop}%
\bibitem [{\citenamefont {Kailasnath}\ \emph {et~al.}(1992)\citenamefont
  {Kailasnath}, \citenamefont {Sreenivasan},\ and\ \citenamefont
  {Stolovitzky}}]{krs92}%
  \BibitemOpen
  \bibfield  {author} {\bibinfo {author} {\bibfnamefont {P.}~\bibnamefont
  {Kailasnath}}, \bibinfo {author} {\bibfnamefont {K.~R.}\ \bibnamefont
  {Sreenivasan}}, \ and\ \bibinfo {author} {\bibfnamefont {G.}~\bibnamefont
  {Stolovitzky}},\ }\href@noop {} {\bibfield  {journal} {\bibinfo  {journal}
  {Phys. Rev. Lett.}\ }\textbf {\bibinfo {volume} {68}},\ \bibinfo {pages}
  {2766} (\bibinfo {year} {1992})}\BibitemShut {NoStop}%
\bibitem [{\citenamefont {Kolmogorov}(1962)}]{K62}%
  \BibitemOpen
  \bibfield  {author} {\bibinfo {author} {\bibfnamefont {A.~N.}\ \bibnamefont
  {Kolmogorov}},\ }\href@noop {} {\bibfield  {journal} {\bibinfo  {journal} {J.
  Fluid Mech.}\ }\textbf {\bibinfo {volume} {13}},\ \bibinfo {pages} {82}
  (\bibinfo {year} {1962})}\BibitemShut {NoStop}%
\bibitem [{\citenamefont {Meneveau}\ and\ \citenamefont
  {Sreenivasan}(1987)}]{menevkrs87}%
  \BibitemOpen
  \bibfield  {author} {\bibinfo {author} {\bibfnamefont {C.}~\bibnamefont
  {Meneveau}}\ and\ \bibinfo {author} {\bibfnamefont {K.~R.}\ \bibnamefont
  {Sreenivasan}},\ }\href@noop {} {\bibfield  {journal} {\bibinfo  {journal}
  {Phys. Rev. Lett.}\ }\textbf {\bibinfo {volume} {59}},\ \bibinfo {pages}
  {1424} (\bibinfo {year} {1987})}\BibitemShut {NoStop}%
\bibitem [{\citenamefont {She}\ and\ \citenamefont {Leveque}(1994)}]{SL94}%
  \BibitemOpen
  \bibfield  {author} {\bibinfo {author} {\bibfnamefont {Z.~S.}\ \bibnamefont
  {She}}\ and\ \bibinfo {author} {\bibfnamefont {E.}~\bibnamefont {Leveque}},\
  }\href@noop {} {\bibfield  {journal} {\bibinfo  {journal} {Phys. Rev. Lett.}\
  }\textbf {\bibinfo {volume} {72}},\ \bibinfo {pages} {336} (\bibinfo {year}
  {1994})}\BibitemShut {NoStop}%
\bibitem [{\citenamefont {Yakhot}(2001)}]{yakhot01}%
  \BibitemOpen
  \bibfield  {author} {\bibinfo {author} {\bibfnamefont {V.}~\bibnamefont
  {Yakhot}},\ }\href@noop {} {\bibfield  {journal} {\bibinfo  {journal} {Phys.
  Rev. E}\ }\textbf {\bibinfo {volume} {63}},\ \bibinfo {pages} {026307}
  (\bibinfo {year} {2001})}\BibitemShut {NoStop}%
\bibitem [{\citenamefont {Migdal}(1995)}]{migdal95}%
  \BibitemOpen
  \bibfield  {author} {\bibinfo {author} {\bibfnamefont {A.~A.}\ \bibnamefont
  {Migdal}},\ }in\ \href@noop {} {\emph {\bibinfo {booktitle} {The first Landau
  Institute summer school, Selected proceedings}}},\ \bibinfo {editor} {edited
  by\ \bibinfo {editor} {\bibfnamefont {V.~P.}\ \bibnamefont {Mineev}}}\
  (\bibinfo  {publisher} {London: Gordon and Breach Publishers},\ \bibinfo
  {year} {1995})\ pp.\ \bibinfo {pages} {177--204}\BibitemShut {NoStop}%
\bibitem [{\citenamefont {Meneveau}\ \emph {et~al.}(1990)\citenamefont
  {Meneveau}, \citenamefont {Sreenivasan}, \citenamefont {Kailasnath},\ and\
  \citenamefont {Fan}}]{MSF90}%
  \BibitemOpen
  \bibfield  {author} {\bibinfo {author} {\bibfnamefont {C.}~\bibnamefont
  {Meneveau}}, \bibinfo {author} {\bibfnamefont {K.~R.}\ \bibnamefont
  {Sreenivasan}}, \bibinfo {author} {\bibfnamefont {P.}~\bibnamefont
  {Kailasnath}}, \ and\ \bibinfo {author} {\bibfnamefont {M.~S.}\ \bibnamefont
  {Fan}},\ }\href@noop {} {\bibfield  {journal} {\bibinfo  {journal} {Phys.
  Rev. A}\ }\textbf {\bibinfo {volume} {41}},\ \bibinfo {pages} {894} (\bibinfo
  {year} {1990})}\BibitemShut {NoStop}%
\bibitem [{\citenamefont {Sreenivasan}(1984)}]{KRS84}%
  \BibitemOpen
  \bibfield  {author} {\bibinfo {author} {\bibfnamefont {K.~R.}\ \bibnamefont
  {Sreenivasan}},\ }\href@noop {} {\bibfield  {journal} {\bibinfo  {journal}
  {Phys. Fluids}\ }\textbf {\bibinfo {volume} {27}},\ \bibinfo {pages} {1048}
  (\bibinfo {year} {1984})}\BibitemShut {NoStop}%
\bibitem [{\citenamefont {Sreenivasan}(1998)}]{KRS98}%
  \BibitemOpen
  \bibfield  {author} {\bibinfo {author} {\bibfnamefont {K.~R.}\ \bibnamefont
  {Sreenivasan}},\ }\href@noop {} {\bibfield  {journal} {\bibinfo  {journal}
  {Phys. Fluids}\ }\textbf {\bibinfo {volume} {10}},\ \bibinfo {pages} {528}
  (\bibinfo {year} {1998})}\BibitemShut {NoStop}%
\bibitem [{\citenamefont {Pearson}\ \emph {et~al.}(2002)\citenamefont
  {Pearson}, \citenamefont {Krogstad},\ and\ \citenamefont {Van
  De~Water}}]{pearson2002}%
  \BibitemOpen
  \bibfield  {author} {\bibinfo {author} {\bibfnamefont {B.}~\bibnamefont
  {Pearson}}, \bibinfo {author} {\bibfnamefont {P.-{\AA}.}\ \bibnamefont
  {Krogstad}}, \ and\ \bibinfo {author} {\bibfnamefont {W.}~\bibnamefont {Van
  De~Water}},\ }\href@noop {} {\bibfield  {journal} {\bibinfo  {journal} {Phys.
  Fluids}\ }\textbf {\bibinfo {volume} {14}},\ \bibinfo {pages} {1288}
  (\bibinfo {year} {2002})}\BibitemShut {NoStop}%
\end{thebibliography}%


%
\end{document}